# Magnet stability and reproducibility

*N. Marks*
ASTeC, STFC, Daresbury Laboratory and University of Liverpool,
The Cockcroft Institute, Daresbury, Warrington WA4 4AD, U.K.

**Abstract**
Magnet stability and reproducibility have become increasingly important as greater precision and beams with smaller dimension are required for research, medical and other purpose. The observed causes of mechanical and electrical instability are introduced and the engineering arrangements needed to minimize these problems discussed; the resulting performance of a state-of-the-art synchrotron source (Diamond) is then presented. The need for orbit feedback to obtain best possible beam stability is briefly introduced, but omitting any details of the necessary technical equipment, which is outside the scope of the presentation.

## 1 Introduction

Magnet stability and reproducibility clearly are important issues that strongly affect the performance of accelerator magnets and influence their effectiveness in interacting with the beam as required. Skilful design and high-quality engineering are worthless if variations of the magnet performance over time (i.e., lack of stability) or between the different elements in lattice (i.e., poor reproducibility) have a deleterious effect on beam properties. Notwithstanding, the quest for ultra-high stability is a relatively recent phenomenon that has appeared with the advent of larger and larger accelerators generating beams with smaller and smaller cross-sections. The subject is also quite specialized, with a relatively small number of individuals practising their expertise in this area; and it spans a wide range of sciences and technologies, for example, from geology and seismology, to mechanical and electrical engineering and, of course, accelerator theory.

For these reasons, the presentation and this summary article rely heavily on published material or private communication from a number of experts in this field, who are acknowledged at the end of the paper.

## 2 Beam stability requirements

Starting at the delivery end of the accelerator complex, the degree of beam stability required at the interaction point (whether on a target, a sample, at a patient, or on another beam), depends very much on the purpose of the accelerator. In this paper the requirements typical of some of the most demanding applications — for example, particle physics research and materials analysis — will be considered. In all facilities, the beam stability is an important component of the accelerator specification, but the areas cited perhaps present the biggest challenges. It is clear that the higher precision demanded by the research community can be met by reducing beam size only if the beam position and angle is correspondingly stable and movement and deflection is restricted to some small fraction of the beam emittance.

The resulting demand for magnet stability and operational reproducibility then depends substantially on the lattice parameters; this issue will be addressed in Section 3, whilst the parameters that drive these requirements — beam geometries and required positional and directional stability — are detailed below, with examples chosen from two existing facilities and a proposed future project.

## 2.1 A ‘state-of-the-art’ synchrotron source — Diamond

The magnet lattices of the electron storage rings, that are the core of a synchrotron source, have evolved significantly since the first dedicated sources were designed with simple FODO structures in the 1970s. The state-of-the-art facilities now have specialized straight sections housing insertion devices, and many quadrupole families producing beams with very small cross-sections. These small dimensions produce very high brightness photon beams for experimenters but, correspondingly, the beam position and angle at each source point needs to be highly stable.

The situation is exemplified by ‘Diamond’ the 3 GeV synchrotron source recently commissioned in the UK. This storage ring has a complex lattice of 48 dipoles and 240 quadrupoles [1], giving the stored electron beam the horizontal and vertical β values shown in Fig. 1.

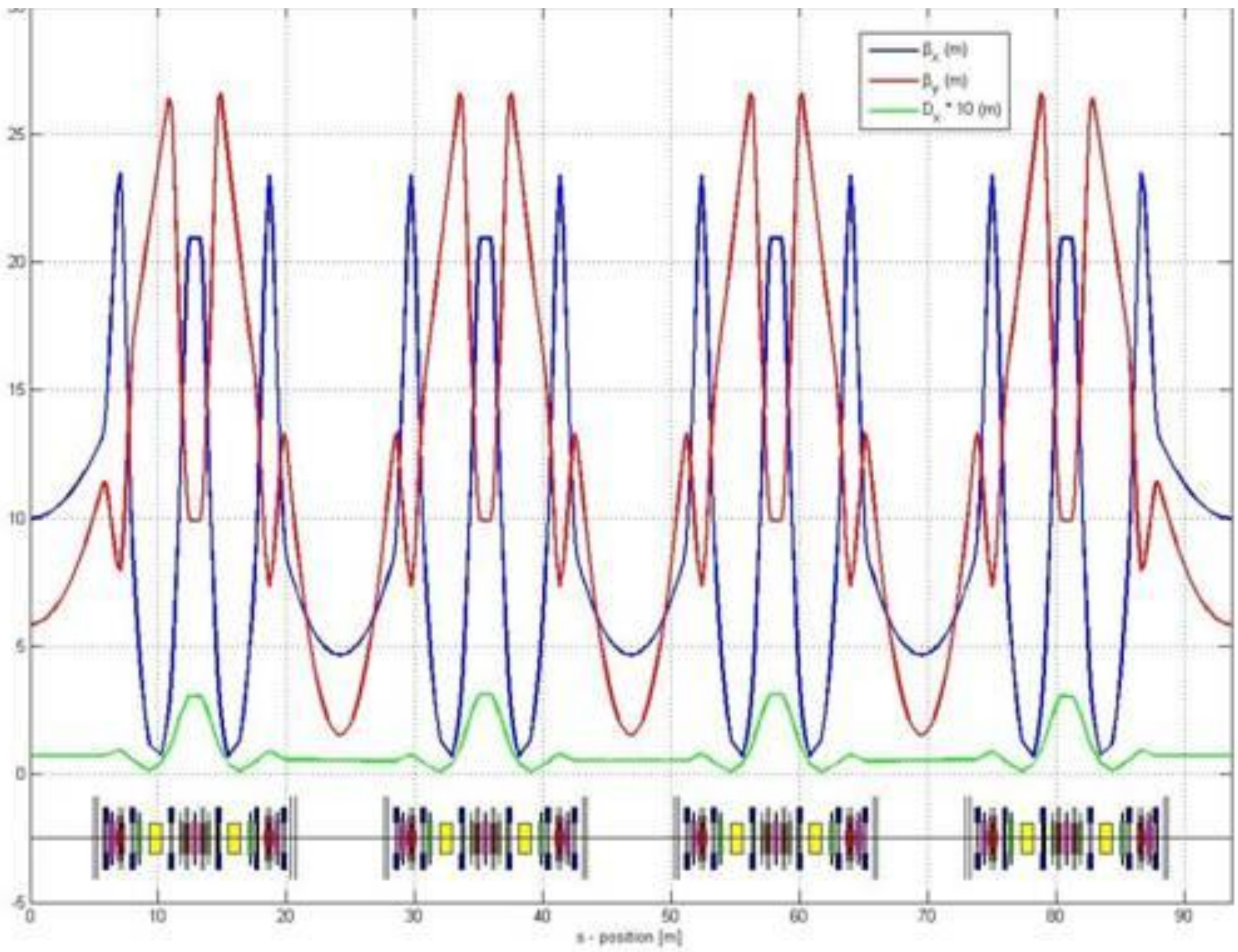

**Fig. 1:** Part of the Diamond magnet lattice showing the curves of $\beta_x$ (blue) and $\beta_y$ (red)

It can be seen that the lattice focuses the beam in both planes in the straight sections; the beam dimensions and divergences at these points are given in Table1, together with the beam emittances in the lattice.

**Table 1:** Electron beam parameters in the third-generation (2005) 3 GeV synchrotron source, Diamond

| | **Horizontal** | **Vertical** |
|---|---|---|
| Beam size fwhh (μm) | 123 | 6.4 |
| Beam divergence (μrad) | 23 | 4.2 |
| Beam emittance (nm rad) | 2.7 | 0.03 |

In the Diamond project, it was determined that the stability of the electron beam at the photon source points needed to be 10% of the beam dimension figures above [2], giving the tolerances on electron beam stabilities given in Table 2.

**Table 2:** Electron beam stabilities required in the 3 GeV synchrotron source, Diamond

| | **Horizontal** | **Vertical** |
|---|---|---|
| Beam positional variation (μm) ≤ | 12.3 | 0.64 |
| Beam angular variation (μrad) ≤ | 2.3 | 0.42 |

### *2.1.1 Comparison with an earlier SR source*

It is of interest to compare the Diamond beam dimensional data with that from the Daresbury SRS, an early dedicated source initially commissioned in the 1970s and upgraded some ten years later to meet the requirements for a second-generation light source. The upgraded lattice gave the electron beam parameters of Table 3.

**Table 3:** Electron beam sizes and emittance in the 2 GeV second-generation (1988) source, SRS

| | **Horizontal** | **Vertical** |
|---|---|---|
| Beam size fwhh (mm) | 2.6 | 0.24 |
| Beam emittance (μm rad) | 0.11 | |

It can be seen that emittances had been reduced by over two orders of magnitude in less than twenty years, with correspondingly increased demands on beam stability. Given the same criterion as applied to Diamond, the tolerance on beam movement in the SRS was of the order of 0.3 mm horizontally and 0.024 mm vertically — values that could be achieved at that time with conventional technology.

## 2.2 The CERN LHC

The Large Hadron Collider (LHC) is, at the time of writing, being commissioned at CERN. The storage ring will collide beams travelling in opposite directions, with beam cross-sections in the interaction regions determining the experimental luminosity. The expected sizes of a 7 TeV beam in the region of the CMS interaction point [3] are shown in Fig. 2. The expected r.m.s. values at a number of collision points are given in Table 4.

**Table 4:** Expected r.m.s. beam sizes at interaction points in the LHC

| | |
|---|---|
| CMS & ATLAS (protons) | 16 μm |
| LHC b (protons) | 22–160 μm |
| ALICE (ions) | 16 μm |
| ALICE (protons) | > 160 μm |

It can be seen that at the narrowest interaction points, the LHC beam will be approximately an order of magnitude smaller than the horizontal electron beam size in Diamond, resulting in a requirement for correspondingly greater stability.

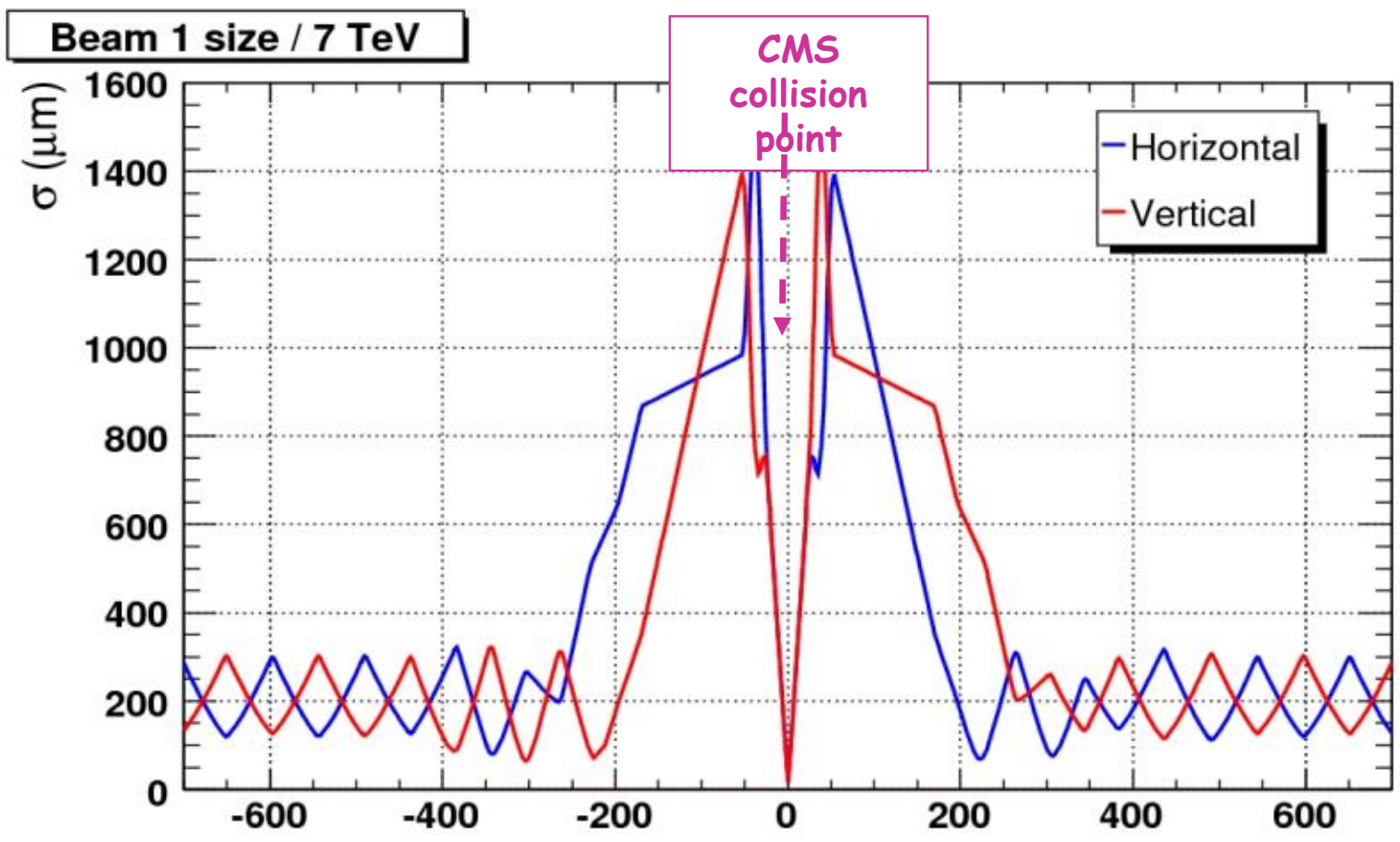

**Fig. 2:** Expected rms beam sizes at 7 TeV at the CMS collision point in the CERN LHC

## 2.3 The proposed International Linear Collider (ILC)

Considerable work has already taken place to design a future particle physics facility that will collide 250 GeV electrons together, with an upgrade path to 500 GeV per beam; a diagram of the proposed layout is given in Fig. 3.

Whilst the project is yet to be funded, the critical issue of the required electron beam dimensions has been addressed [4]. The cited presentation explains:

*'After acceleration, the beams are not ready to deliver the full luminosity required for the physics studies – the beam size must be reduced from 1 micrometre at the end of the acceleration unit to just a few nanometres at the interaction point …. (they) need to go through several optics correction points….'*

The presentation does not indicate the tolerance that this microscopic beam places on positional and angular stability but it is clear that any movement of the same order as the total beam size would inhibit the collision process, destroy the luminosity, and therefore would be unacceptably large.

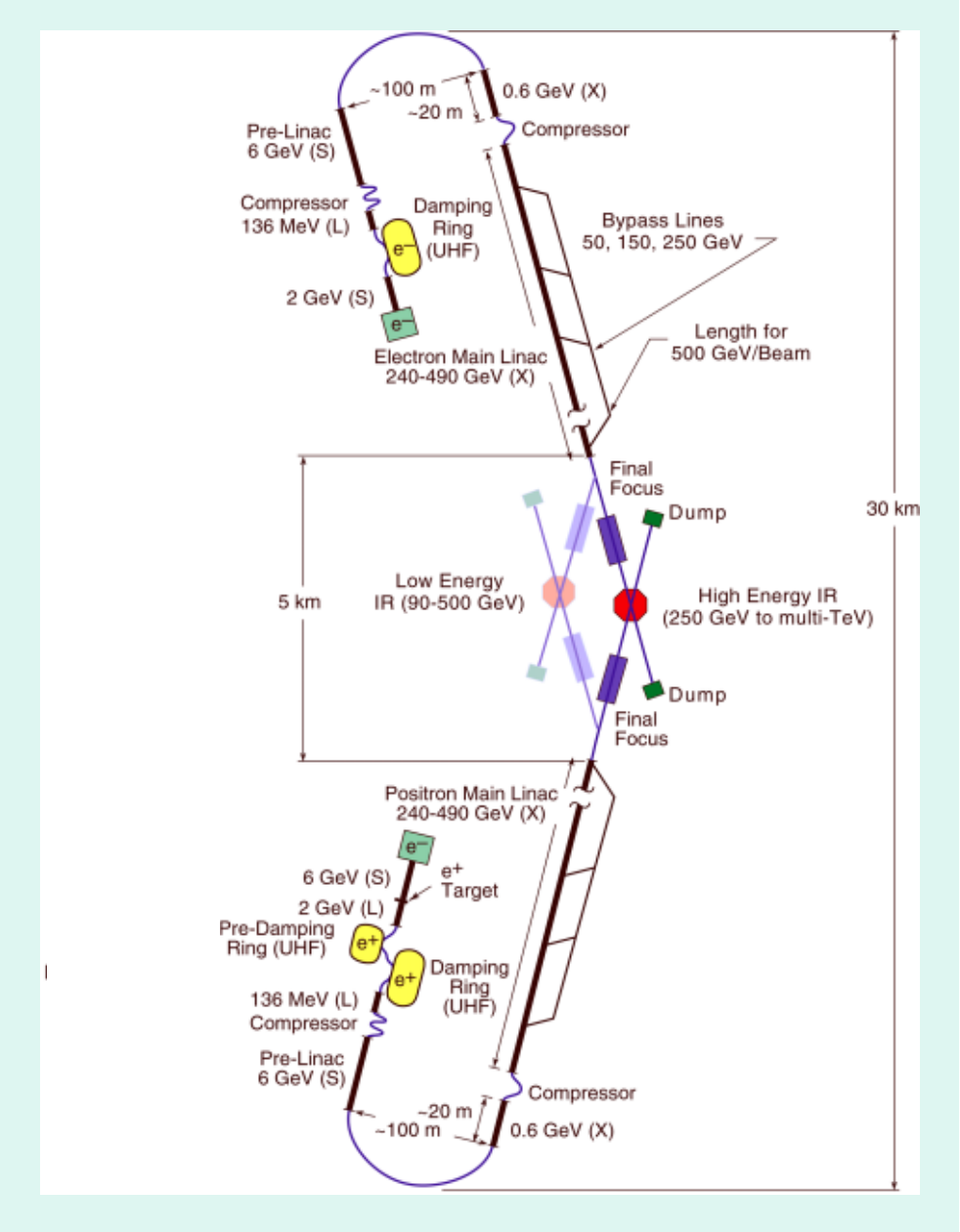

**Fig. 3:** Proposed layout of the 30 km long e- / e+ International Linear Collider (ILC)

The implication from these three examples is that the most demanding current facilities require beam stabilities of the order of a few microns in the horizontal plane and less than one micron in the vertical. In the future, these tolerances will become significantly tighter, with beam dimensions reducing to the nanometre level.

The implication of these beam requirements on magnet stabilities will now be considered.

# 3 Magnet stabilities

In this section, the effect of magnet positional and angular vibration (due to instabilities) and static displacement (reproducibility factors) on beam position will be considered. The discussion will mainly be concerned with the consequences of such movements of dipoles and quadrupoles, with sextupoles briefly considered.

## 3.1 Dipole magnets

In a circular accelerator, the dipoles generate the vertical field needed to deflect the beam and generate a quasi-circular complete 360˚ orbit; they are therefore the principal magnets present in the accelerator ring. In many accelerators, pure dipole fields are required to satisfy this bending role, the beam focusing being provided by separate quadrupole magnets (a ‘separated-function’ lattice). However, in a minority, a spatial transverse gradient is added to provide some focusing as well as bending, producing a ‘combined-function’ magnet. The issues dealt with in this section apply only to magnets generating a pure dipole field; where combined-function magnets are being used, the stability criteria relevant to quadrupoles, as described in the next section, will also apply and must be duly considered.

Whilst these issues are discussed in the context of a circular accelerator, they apply equally to beam-lines within which the charged particles are transported in a single pass.

### *3.1.1 Positional and angular displacements*

– Small horizontal and vertical transverse displacements: the dipole design is intended to generate a uniform vertical field over the ‘good field’ aperture of the magnet in both the horizontal and vertical planes. Thus, with any small transverse displacement, the vertical field at the beam remains unchanged and hence has a negligible or zero effect on the beam; large displacement would result in some part of the beam path being outside the good field region of the magnet, but such movements are gross and very much larger than the instabilities that are considered.

– Longitudinal displacement: this also does not affect the integrated filed strength of the magnet experienced by the beam and therefore the dipole still will generate the correct beam angular bend; however, it changes the position of the azimuthal centre of that bend in the lattice, creating a loss of symmetry in the complete circular ring; this will produce some closed-orbit distortion (i.e., horizontal displacement) of the beam around the complete ring; the degree of such distortion will completely depend on the lattice and needs to be examined for each particular design; it should also be noted that if the displacement is static (i.e., a positional reproducibility error), small distortions can be corrected by the beam-steering d.c. correctors, whilst longitudinal vibration (instability) could be expected to be sufficiently small to generate no appreciable distortion; notwithstanding, good communication between magnet engineers and lattice designers will be needed to resolve this issue and it is likely that a more stringent tolerance would be defined on longitudinal dipole placement than in the two transverse planes.

– Twist about the longitudinal axis (‘roll’ — see Fig. 4 for definition of angular errors); roll is a far more serious issue; a roll error of θ in a dipole will generate a small horizontal field:

$$B_x = B_y \sin \theta;$$

the horizontal field $B_x$ will produce a vertical bend in the beam and, whilst $\theta$ will be very small, the main bending field $B_y$ will be large and an unacceptable vertical steering effect may be produced, significantly distorting the orbit in the vertical plane; the principal issue will be static — the accuracy with which the magnet is surveyed into position — but dynamic effects (vibration) must also be carefully considered; an error that must be minimized, with a tolerance band to be specified by the lattice designers.

– Twist about the radial axis ('pitch'): this will produce an axial field component which will change horizontal and vertical focusing and will couple horizontal and vertical betatron oscillations; this is an undesirable effect in any accelerator but a critical issue in high-energy lepton machines where the vertical beam size is determined principally by this coupling effect; again, this error must be minimized to meet lattice specifications.

– Twist about a vertical axis ('yaw'): possibly less critical than roll or pitch, as it does not couple motion in the two transverse planes; it does produce an entry and exit angle at each end of the magnet which will result in some transverse focusing effects which will, in principle, cancel each other out; a further topic for consultation with lattice experts.

–

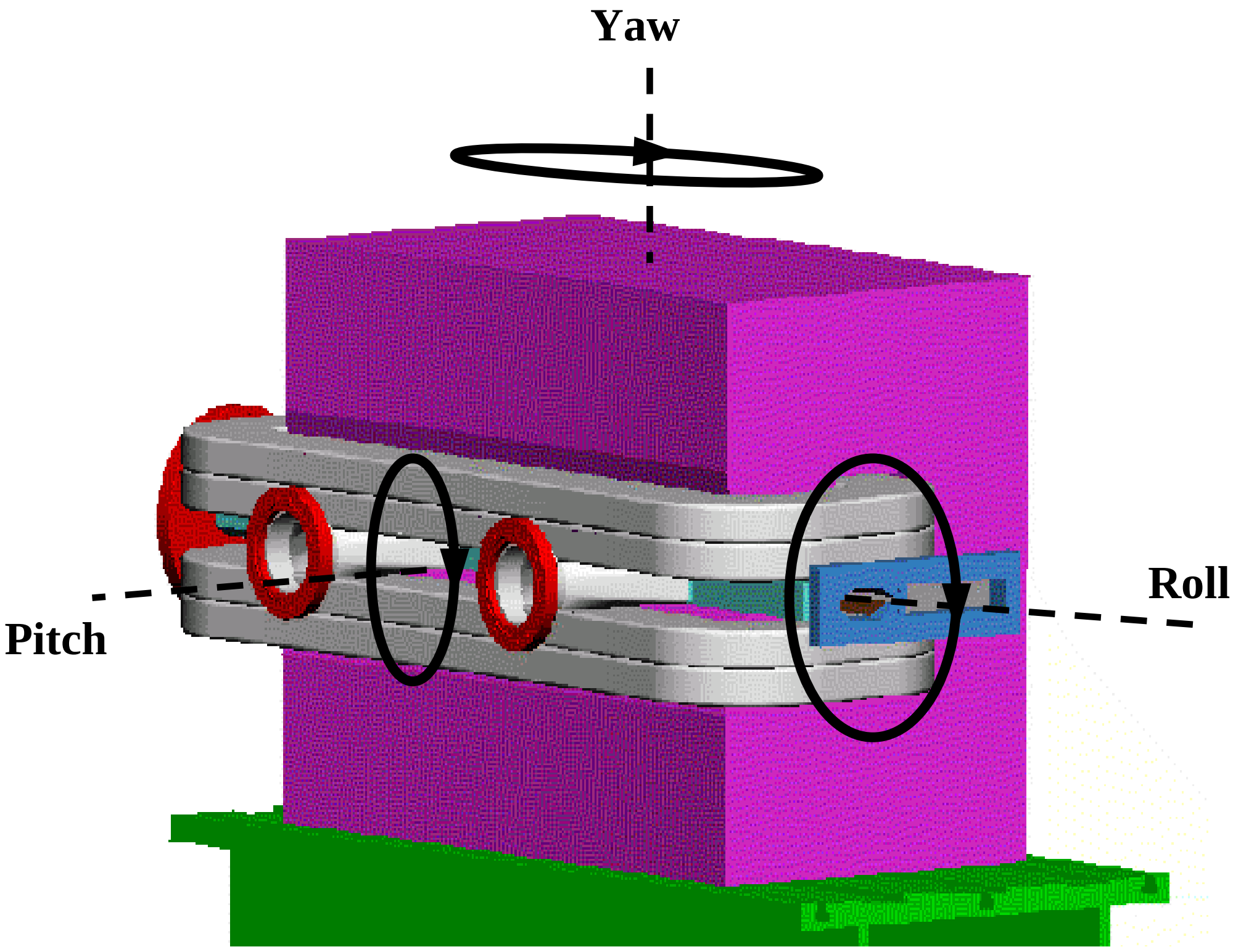

**Fig. 4:** Definition of terms used to describe angular errors in magnet alignment

### *3.1.2 Dipole strength variations*

The beam deflection produced by a dipole magnet is proportional to its 'magnetic strengths' — the azimuthal integration of the vertical field through the length of the magnet:

$$\int B_y.dz$$

Variations in strength are produced by:

- errors in reproducibility in magnet geometry, in both the inter-pole gap (or coil configuration for a s.c. magnet) and physical length, occurring during manufacture; these can be minimized by choosing the lattice position for the dipoles once their magnetic strengths have been measured after production;
- current leakage from the circuit powering the magnets' coils;
- where dipole string is separated into separate circuits, inequality of the output current from different power converters.

It is this parameter that is one of the most critical in a circular accelerator, for variations in bend-strengths within a lattice made up of (nominally identical) separate dipoles, will induce major orbit distortions. Again, the tolerance that can be placed on an acceptable variation in strength, dipole to dipole, depends critically on the lattice details but it is possible to give some general values for accelerators of different sizes. A tolerance of $\pm$ 1:$10^4$ is typical for smaller machines, with greater reproducibility being required for larger installations; for example, the high-current dipole power converters for the LHC require a d.c. stability of better than 10 ppm [5].

## 3.2 Quadrupole magnets

### *3.2.1 Quadrupole transverse positional variation*

Quadrupole magnets are required to generate zero field at the correct beam centre, with a linear field gradient across the beam, in both horizontal and vertical planes, to focus particles that are off-centre. The zero field point (the 'magnetic-centre' where $B_x = B_y = 0$) therefore needs to be accurately located on the correct, undeviated closed orbit. Any subsequent transverse displacement of the magnet will consequently result in the deflection of the central closed-orbit, moving the beam position in horizontal or vertical position (depending on the direction of the magnet displacement) at all other parts of the ring.

The amplitude of such beam displacement depends critically on the lattice design but, in a strong-focusing configuration, it will be very much greater than the magnet displacement that is causing the orbit distortion. The ratio of beam movement to quadrupole displacement is known as the '**amplification factor**'. For a simple FODO lattice (such as the Daresbury SRS) this factor is between 10 and 20; for a modern, complex lattice with low-beta insertion points, it is significantly larger. In the Diamond facility, for example, the amplification factors in the storage ring are

**horizontal amplification factor: 60;**

**vertical amplification factor: 45.**

Hence, transverse displacements in quadrupole magnets produce large orbit distortions and therefore are critical issues in determining beam stability. Continuing with the Diamond facility as an example and taking the figures for the required **beam** stability given in Table 2, the corresponding **quadrupole magnet** stabilities are as shown in Table 5:

**Table 5:** Quadrupole magnet transverse stabilities required in the 3 GeV synchrotron source, Diamond

| | **Horizontal** | **Vertical** |
|---|---|---|
| Quadrupole transverse positional variation (μm) ≤ | 0.2 | 0.015 |

This clearly is a very demanding stability requirement and it is reasonable to question whether a positional stability of 0.015 μm in the vertical plane is achievable. This will be resolved in the later stages of this paper.

#### *3.2.2 Variation in quadrupole strength*

Static variation of strength from quadrupole to quadrupole, due to small errors in engineering reproducibility during construction and assembly of the magnets, will distort the beta values around the lattice and lead to different beam sizes at different circumferential positions around a circular machine. As with other instabilities, the amplitude of this distortion is dependent on the lattice and the acceptable level on the accelerator application. These effects must be estimated and the level of inter-quadrupole strength variation considered during magnet construction. Strength measurements on the complete quadrupole production sequence will indicate whether this target has been achieved, and adjustment of the quadrupole magnetic lengths by small modification at the magnet ends can be carried out if necessary.

Dynamic variations in strength of individual quadrupoles, caused by variation in magnet excitation of the separate power sources for those quadrupoles, are more serious and will lead to instability in beam dimensions at different parts of the ring. This is controlled by ensuring that the specification for the power converters provides for sufficient stability in output current. It should be noted that in some circumstances beta variations around a lattice are required (in a light source with different insertion devices in the various straights for example) and separate power sources are then essential, with the potential of varying their output currents individually.

Dynamic variations in strength of a complete quadrupole family will result in changes of the beam's 'tunes' — the number of betatron oscillations per revolution — in both the radial and vertical planes. This is highly undesirable, for it can cause the beam to engage a resonance resulting in beam transverse disturbances and loss. It must be minimized to an acceptable level, again by imposing a tight tolerance on the stability of the quadrupole power supplies and ensuring that the equipment meets this specification.

### 3.3 Summary of instability effects

The conclusion of the above discussion is that magnet positional and amplitude stability is important in all accelerators and critical in many. It should be borne in mind however, that:

- static positional displacements are corrected during installation and survey;
- static amplitude variations should be corrected by measurement at the end of manufacture;
- uniform positional displacement of ALL accelerator magnets by the same amount in the same direction is not a problem.

So independent (between magnet and magnet), dynamic (time varying) instabilities are of major concern, with quadrupole position being the most critical issue and presenting the biggest problems.

## 4 The causes of magnet-induced beam instabilities

Having examined

- the typical beam stabilities required in various types of facility,
- the relationship between the variations in magnet strengths and positions and the beam instabilities that they cause,
- the resulting magnet positional tolerances required in a modern state-of-the art facility,

we now need to explore further the root causes of the magnet instabilities, so that effective counter-measures can be considered.

An overview of this topic was presented at the 2009 Particle Accelerator Conference in Vancouver [6]. The speaker detailed the causes of positional variation as a function of disturbance frequency, as measured in the ALS at the Lawrence Berkeley National Laboratory, shown in Fig. 5.

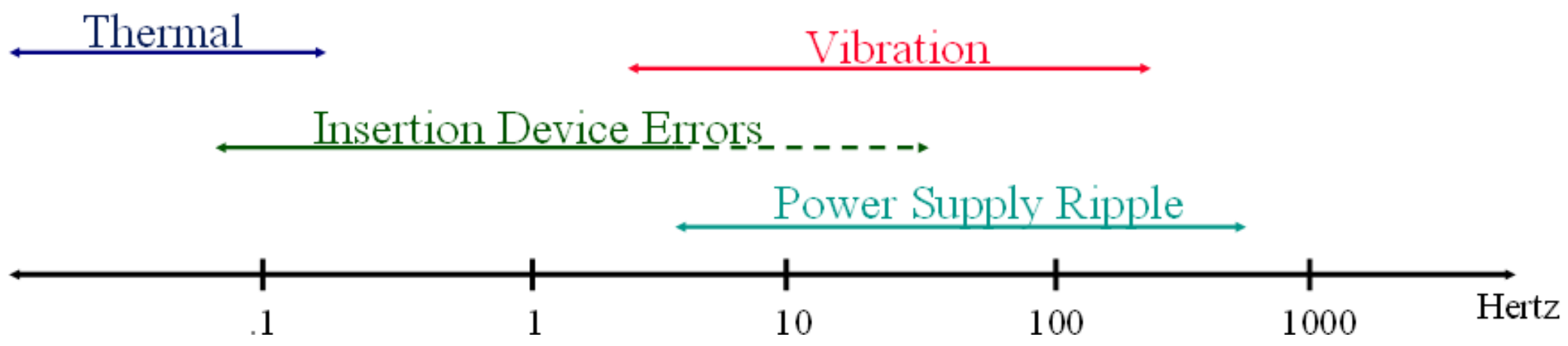

**Fig. 5:** Spectrum of the causes of magnet instability seen in the ALS (LBNL)

The paper also gave further details of the amplitudes of the disturbances and their dominant causes, at different frequencies, as shown in Table 6 below.

**Table 6:** The amplitudes and dominant causes of instabilities perceived in the ALS (LBNL)

| Frequency | Magnitude | Dominant Cause |
|---|---|---|
| Two weeks (A typical experimental run) | ±200 μm Horizontal<br>±100 μm Vertical | 1. Magnet hysteresis<br>2. Temperature fluctuations<br>3. Component heating between 1.5 GeV and 1.9 GeV |
| 1 Day | ±125 μm Horizontal<br>±50 μm Vertical | Temperature fluctuations |
| 8 Hour Fill | ±50 μm Horizontal<br>±20 μm Vertical | 1. Temperature fluctuations<br>2. Feed forward errors |
| Minutes | 1 to 5 μm | 1. Feed forward errors<br>2. D/A converter digitization noise |
| .1 to 300 Hz | 3 μm Horizontal<br>1 μm Vertical | 1. Ground vibrations<br>2. Cooling water vibrations<br>3. Power supply ripple<br>4. Feed forward errors |

Based upon these data, the following sources of magnet instability will be examined and, bearing in mind the discrepancy between the above figures and the required stabilities given in Table 5, possible counter-measures discussed:

- ground vibration;
- thermal instabilities;
- water vibration;
- power supply instabilities and ripple.

## 4.1 Ground vibration

Seismologists and surveyors express ground vibrations measurements as the '**power spectral density**' (PSD), which has dimensions of $(\text{length})^2\,(\text{frequency})^{-1}$ ; in practical circumstances, units of $(\mu\text{m})^2\,(\text{Hz})^{-1}$ are often used. For a particular site, the PSD is measured as a function of frequency and the spectra plotted for the horizontal and vertical planes. The r.m.s value of the physical movement of the ground ($z_{rms}$) is then obtained by taking the square-root of the integrated spectrum ($S_x$) between defined frequency limits ($f_1$ *and* $f_2$):

$$z_{rms}(f_1, f_2) = \sqrt{\int_{f_1}^{f_2} S_x(f)\,df}\ .$$

### *4.1.1 The PSD spectrum*

A typical power spectral density plot is shown in Fig. 6; this was obtained from bed-rock on the Daresbury Laboratory site [7] but has details that are common in all such measurements.

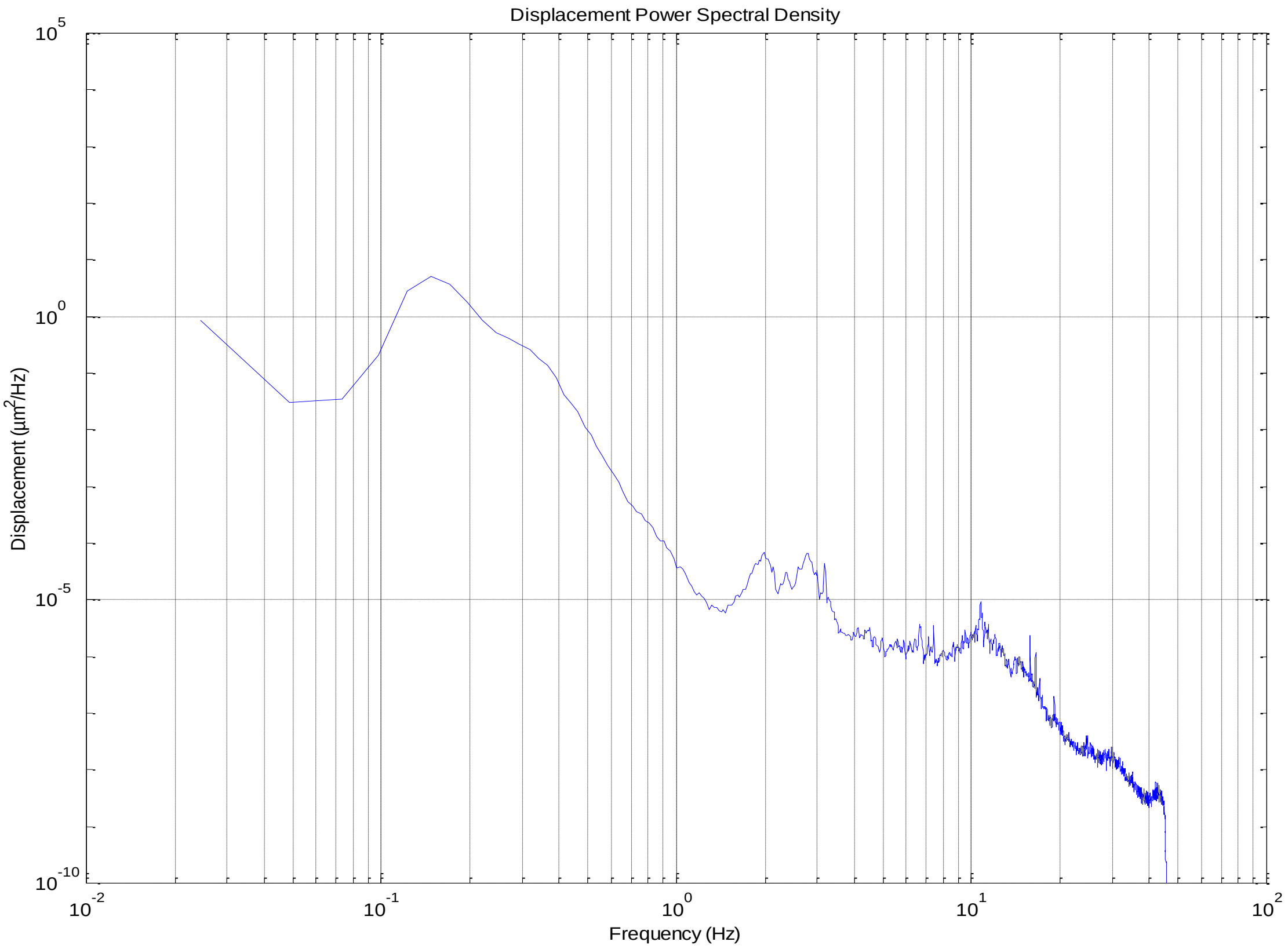

**Fig. 6:** Ground power spectral density in $(\mu\text{m})^2(\text{Hz})^{-1}$ as a function of frequency (Hz), as measured on bed-rock below the Daresbury Laboratory [7]; the axes are logarithmic; the ordinate varies between $10^{-10}$ and $10^5$; the abscissa between $10^{-2}$ and $10^2$.

The peak between 0.1 and 0.3 Hz is called the 'microseismic peak' and is caused by the pounding of ocean waves on the coast-line; it is present on all sites, however far they may be from the sea!

The higher frequencies, at 1 Hz and above, are 'technical and cultural noise'. The amplitude of these disturbances varies significantly between sites, depending on the locality. The proximity of heavy industrial plant, railway tracks or high speed roads will increase this noise, and a laboratory site will also generate its own 'cultural' noise. It is a point of some surprise to learn that an overflying, wide-body aeroplane will depress the ground beneath by up to 4 μm.

The ultra low-frequency 'earth-tide' is semi-diurnal (*c*. $2 \times 10^{-5}$ Hz) and is large: *c*.0.6 m peak to peak, though, owing to its very long wavelength, this does not cause problems.

#### *4.1.2 Ground motion wavelength*

To disturb the beam, magnet vibrations need to be of different amplitudes over the circumference of an accelerator i.e., to be incoherent. The degree of incoherence due to ground motion depends completely on the wavelength of the disturbance. Seismologists have shown that ground disturbances can be categorized into bulk waves and surface waves; there are two types of each and they have different wave velocities and different wavelengths, details depending crucially on the underlying sub-soil and bed-rock formation.

However, Holder [7], when studying the effects of ground vibration on the stability of the lattice magnets in the Diamond storage ring with a 150 m diameter, concludes that 'ground waves with wavelengths of significantly greater than 300 m will not be a problem' and consequently 'the low frequency limit, below which the lattice will move coherently, is about 1.5 Hz'. Clearly, the value of this lower limit depends on the size of the accelerator, being higher for small accelerators, but substantially lower for large machines. He then proceeds to warn that 'above this limit particularly important frequencies exist that give ground wavelengths that are the same order as the betatron wavelengths and therefore cause resonant beam excitations'.

The message is therefore clear — for all but the largest accelerators, the large microseismic peak does not constitute a significant problem, but great attention must be paid to the effects generated by the technical and cultural noise at frequencies of the order of a few Hz and higher. It should be noted from Fig. 6 that there is a plateau in the PSD spectrum between approximately 1 Hz and 10 Hz. Above that the amplitude is decreasing rapidly — approximately three orders of magnitude between 10 Hz and 50 Hz. Whilst the magnitude of the technical and cultural noise will vary from site to site, this rapid decrease is a standard feature. Hence, excitations in the region of 1-10 Hz have the potential to be very damaging.

#### *4.1.3 Girder resonance*

The lattice magnets will be mounted on girders to provide as rigid a support as possible and to facilitate positional adjustments. An engineering FEA model of a Diamond girder, supporting a dipole, four quadrupoles, and three sextupoles on the single girder, is shown in Fig. 7. Ground vibration will be transmitted to the magnets through the girder mounts, so these mechanical engineering components feature strongly in the understanding of the effect of ground movement on the accelerator.

The girders will have a number of resonant frequencies, corresponding to the normal modes of oscillation that can exist in their geometric structure. Steier [6] explains that in the design of some earlier accelerators, massive support girders were used, resulting in low resonant frequencies that were in the 'danger zone' between 1 Hz and 10 Hz. Later, lighter girders were designed but these were still weighted down by mechanical components intended to provide the precision adjustment of magnet position, with the consequence of the resonant frequencies still being low. However, he adds that the latest synchrotron radiation storage rings (Diamond, Soleil and NSLS II) avoid this problem and have successfully 'lifted' the girder resonance frequencies into the tens of hertz.

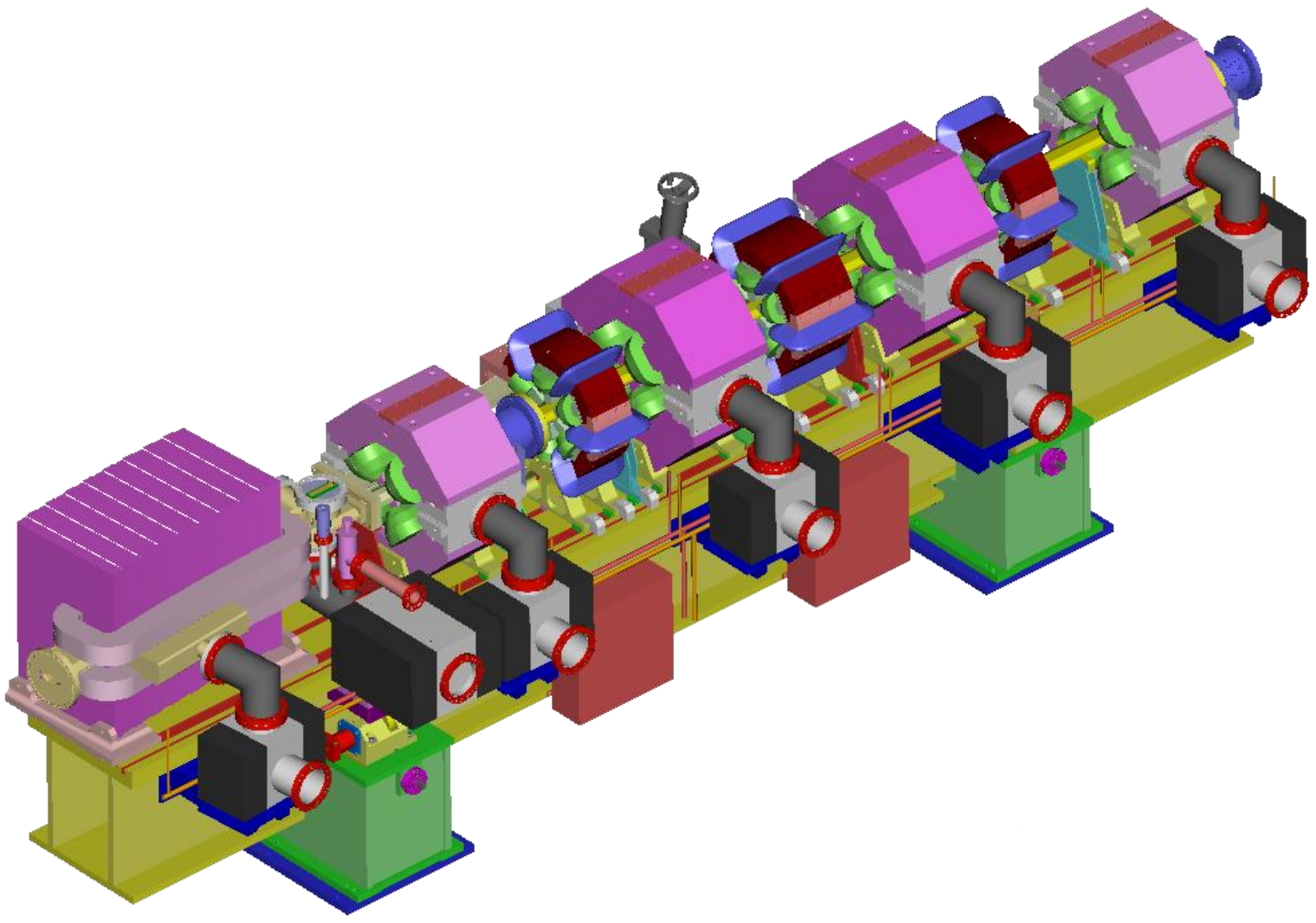

**Fig. 7:** Engineering model of a Diamond girder supporting a dipole, four quadrupoles, and three sextupoles

Huang [8] presents data on the design of the Diamond lattice girders, firstly studying their static deflections. He indicates that the 561 m storage ring uses 72 magnet support girders level between 2 planes, 1 mm apart; the average height difference between adjacent girders is approximately 0.1 mm with a predicted annual variation in level of approximately 0.4 mm. Using finite element analysis (FEA) techniques he predicts the vertical static deflection of the magnets on the girders shown in Fig. 8; the calculated maximum static deflection is 48 µm.

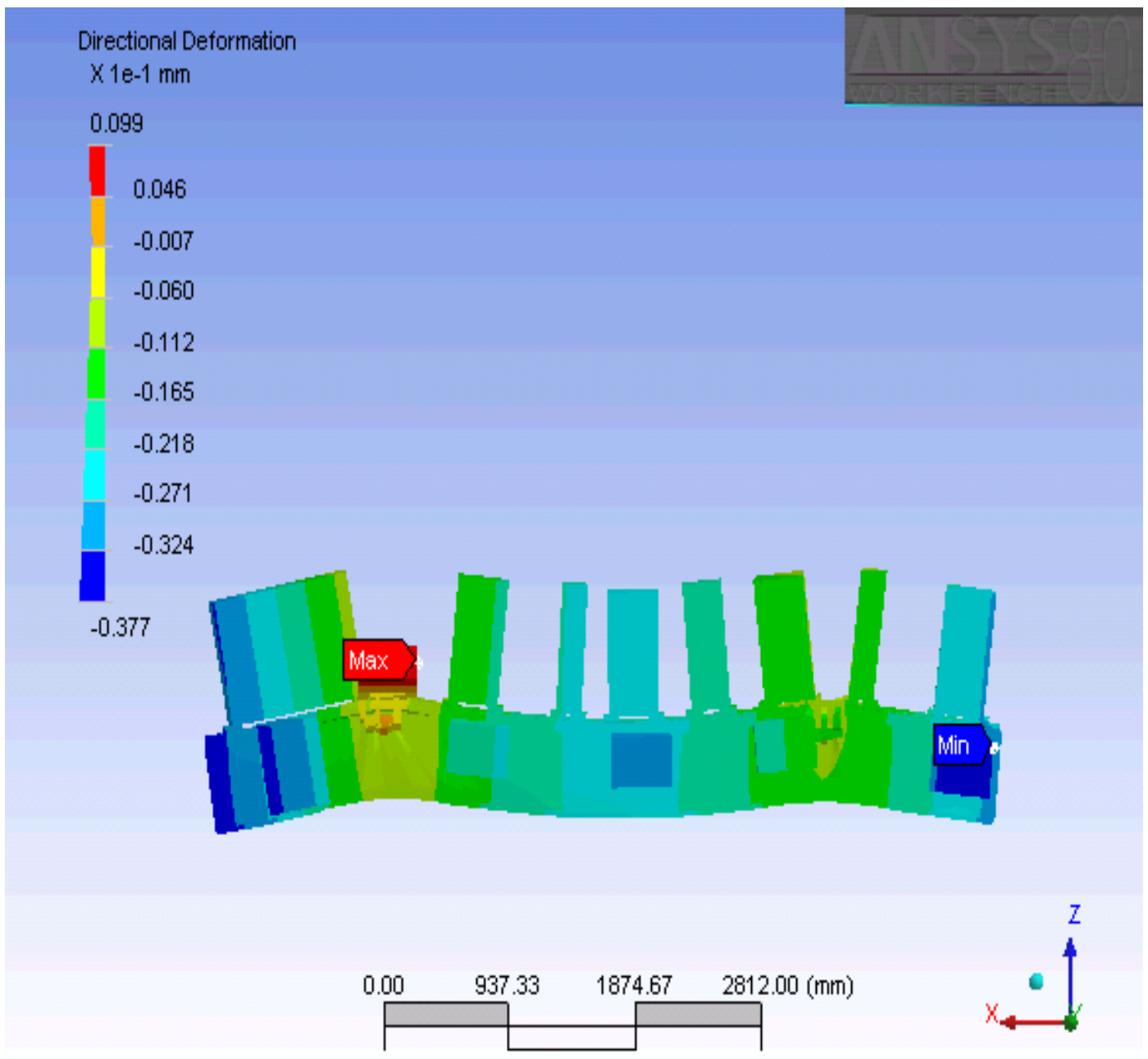

**Fig. 8:** Predicted static vertical deflections of the Diamond lattice girders when loaded with magnets, according to Huang [8]

Clearly, these deflections will need to be taken into account when the magnets are being surveyed into position.

The vertical and horizontal resonance spectra of the girders were also studied using the FEA codes and, knowing the ground power density spectrum on the Diamond site, the consequential spectrum and amplitude of vibration for eight different magnets, positioned on the girders, predicted. The resultant power spectral densities (in units of $mm^2/Hz$) in the vertical and horizontal planes are shown in Figs. 9 and 10.

The predicted vertical resonances are at 41, 51, 53, 63, 73, and 88 Hz, with the horizontal resonance spectrum being similar. Clearly, the resonance frequencies have been raised well above the 10 Hz target minimum.

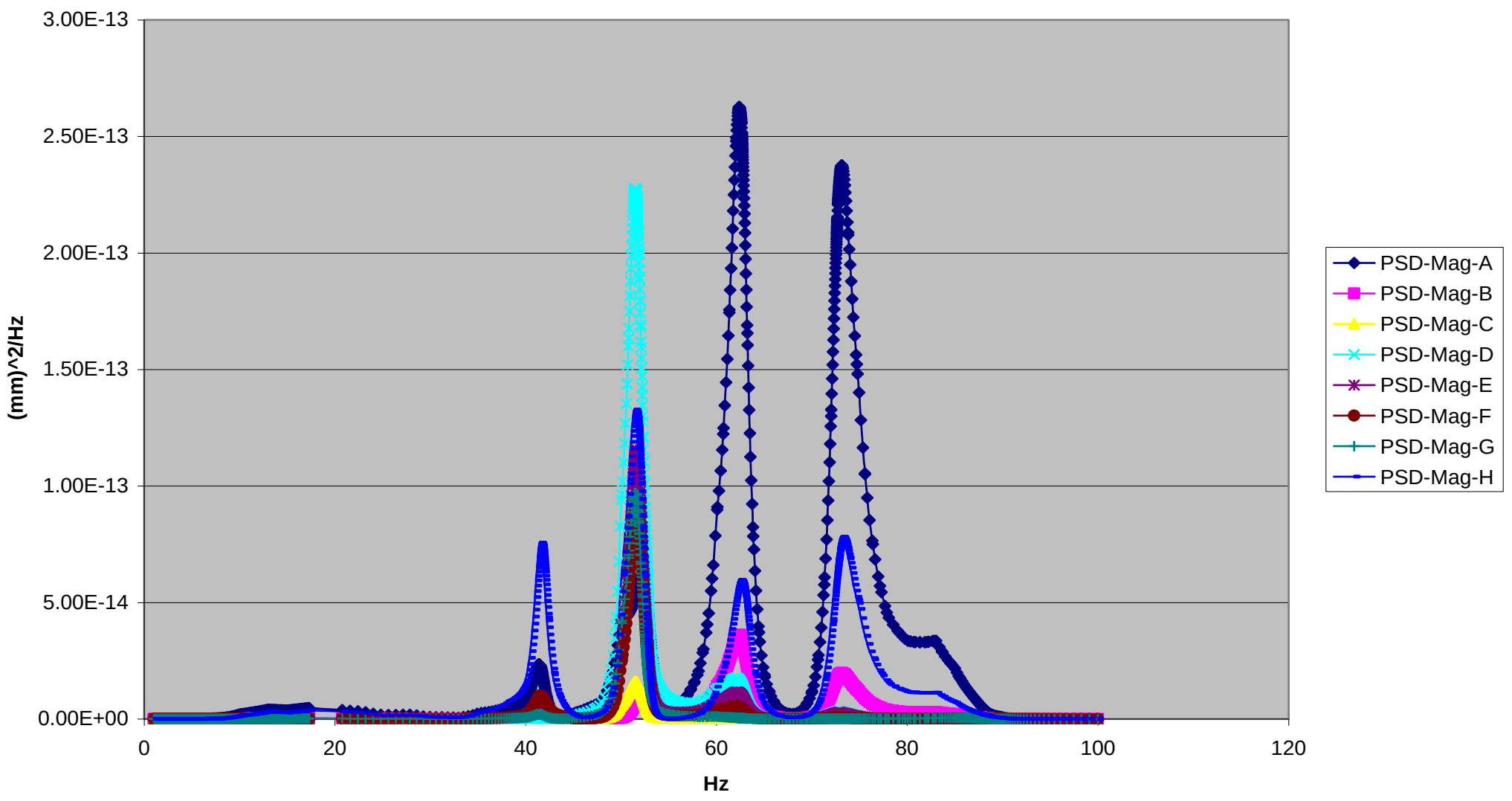

**Fig. 9:** Predicted spectrum of vertical vibrations at the positions of eight magnets in the Diamond lattice, according to Huang [8]; the amplitudes are expressed as the 'power spectral density' in units of $(mm)^2/Hz$.

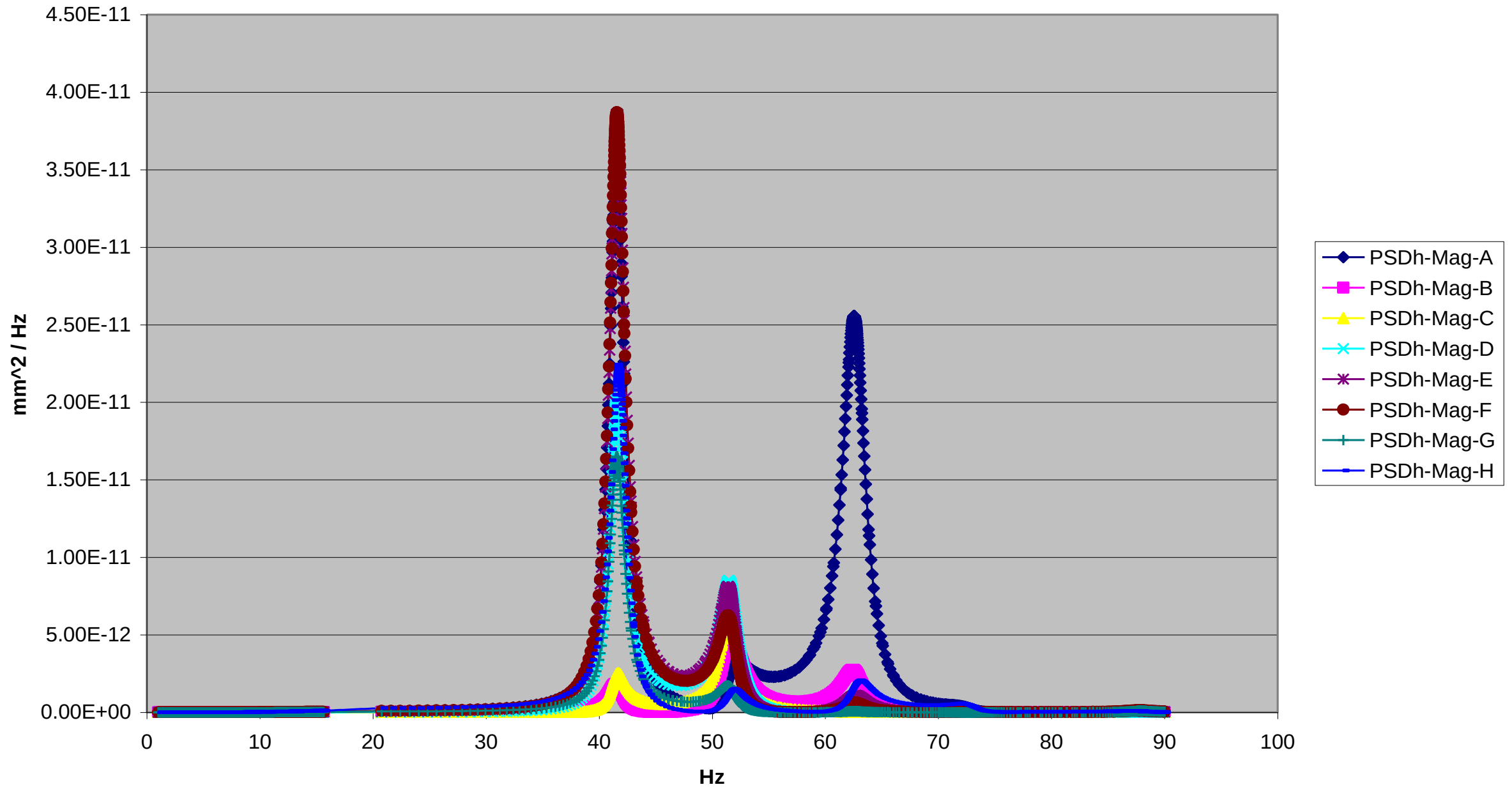

**Fig. 10:** Predicted spectrum of horizontal vibrations at the positions of eight magnets in the Diamond lattice, according to Huang [8]; the amplitudes are expressed as the 'power spectral density' in units of $(mm)^2/Hz$.

## 4.2 Thermal instabilities

With the linear coefficient of expansion of steel being *c*.$1.3 \times 10^{-5}$ /K, the effects of temperature change on magnet strength and position are illustrated by the following example:

A quadrupole with:

inscribed radius of 40 mm;
pole length of 100 mm;
magnetic centre 200 mm above support feet;

then:

strength change due to pole expansion = 0.007% / K;
movement of magnetic centre = 2.5 μm/ K.

So during power-up, with a potential temperature increase of the magnet yoke of up to 10 K, the quadrupole centre would move by up to 25 μm — far in excess of the tolerances required as given in Table 5 — and the strength increase by nearly 0.1% — also unacceptably large.

It is clearly crucial to control the temperature of the accelerator environment — both the air temperature in the tunnel and the water circulated for equipment cooling. Also, the temperature rise in the magnets themselves must be minimized.

### *4.2.1 Ambient temperature control*

The degree of temperature control now economically possible in the large enclosed space of an accelerator lattice tunnel and the circulated cooling water is indicted by Steier [6] by the data presented in Fig. 11, which shows the temperature of the cooling water (left-hand graphs) and tunnel air temperature (right-hand graphs) at the ALS over 24 hours.

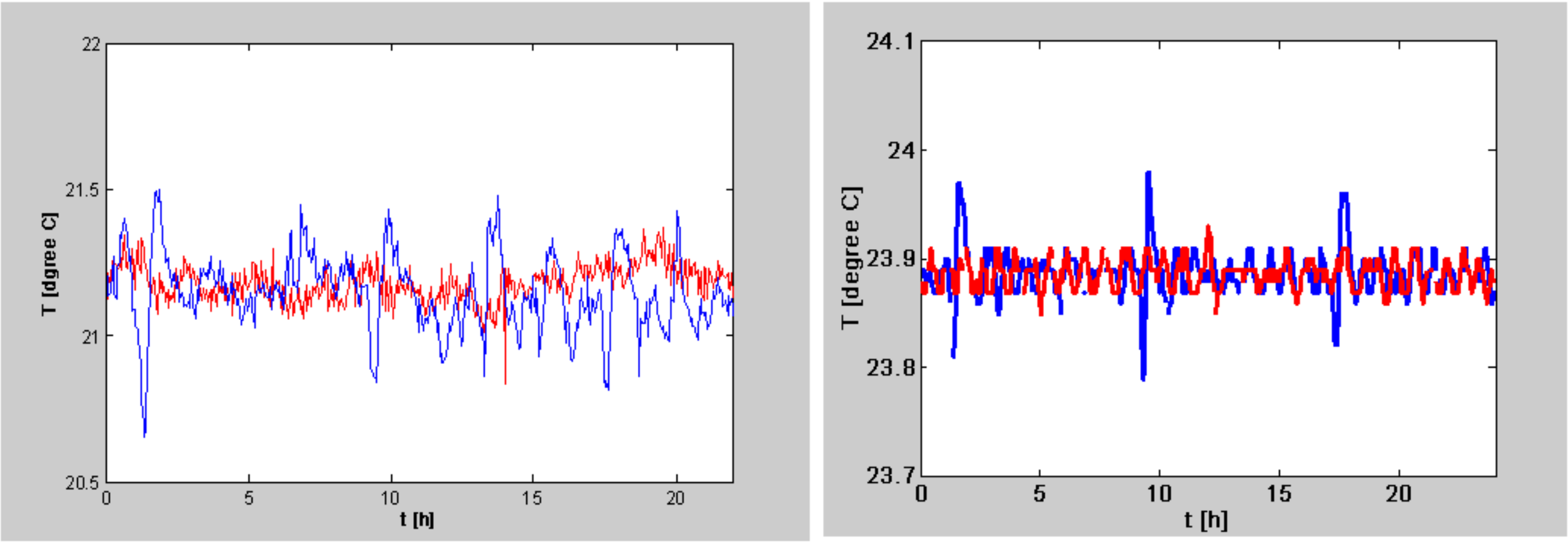

**Fig. 11:** Temperature of the ring cooling water (left-hand ) and tunnel air temperature (right-hand) at the ALS over 24 hours; the red curves are with the tunnel room removed.

The mean and maximum temperature variations in the Diamond storage rings are provided by Kay [9]:

– tunnel air temperature: 22.0 ± 0.1˚C;
– demineralized water supply temperature: 22 ± 0.3˚C.

### *4.2.2 Magnet temperature rise*

However stable the ambient air temperature and the cooling water input temperature, the magnet yokes will run above that value. It is therefore important to limit the temperature rise in the magnet coils, as heat conducted into the yoke will result in dimensional changes to the yoke with corresponding movement to the magnetic centre. The power dissipation in the coils is governed by the root mean square (r.m.s.) of the coil current and the copper cross-section of the total coil.

The r.m.s. current depends on the waveform of the coil current; in the case of a storage ring, where, during operation, the magnets are kept at a fixed dc current, the r.m.s. current is obviously equal to the d.c. value. It is also standard knowledge that for a pure alternating waveform, the r.m.s. current is the peak divided by the square root of 2. But in a booster synchrotron, where the magnets are excited by a monodirectional waveform that comprises a d.c. and an a.c. component, the r.m.s. value is given by the following:

$$I_{rms} = \sqrt{\{I_{dc}^2 + (I_{ac}^2)/2\}}\ .$$

In the particular case of a fully biased sin-wave,

$$I_{dc} = I_{ac} = (1/2)\ I_{peak}$$

and

$$I_{rms} = I_{dc}\ (\sqrt{(3/2)} = I_{peak}\ (1/2)\{\sqrt{(3/2)}\}\ ,$$

where $I_{ac}$ is the peak of the a.c. component and $I_{peak}$ is the current maximum in the waveform.

The choice of conductor cross-section in the coils is determined by economic criteria. The ampere-turns needed are fixed by the magnet geometry and required field value. There is a high value of current density at which cooling becomes difficult, but it is usual to operate well below such a critical value. Examining the economics of building and then operating the magnet systems gives the following conflicting requirements:

– to minimize the capital cost of the coils and yoke, and hence of the whole magnet, as small an amount of conductor as possible should be chosen;

– but this then gives a higher current density, increased power losses, and expensive power expenditure, so lifetime operational costs increase.

Hence, it is usual to consider the total lifetime cost of the magnet system, both construction and operational costs, and to determine the optimum value of current density that minimizes total expenditure. The situation is demonstrated in Fig. 12.

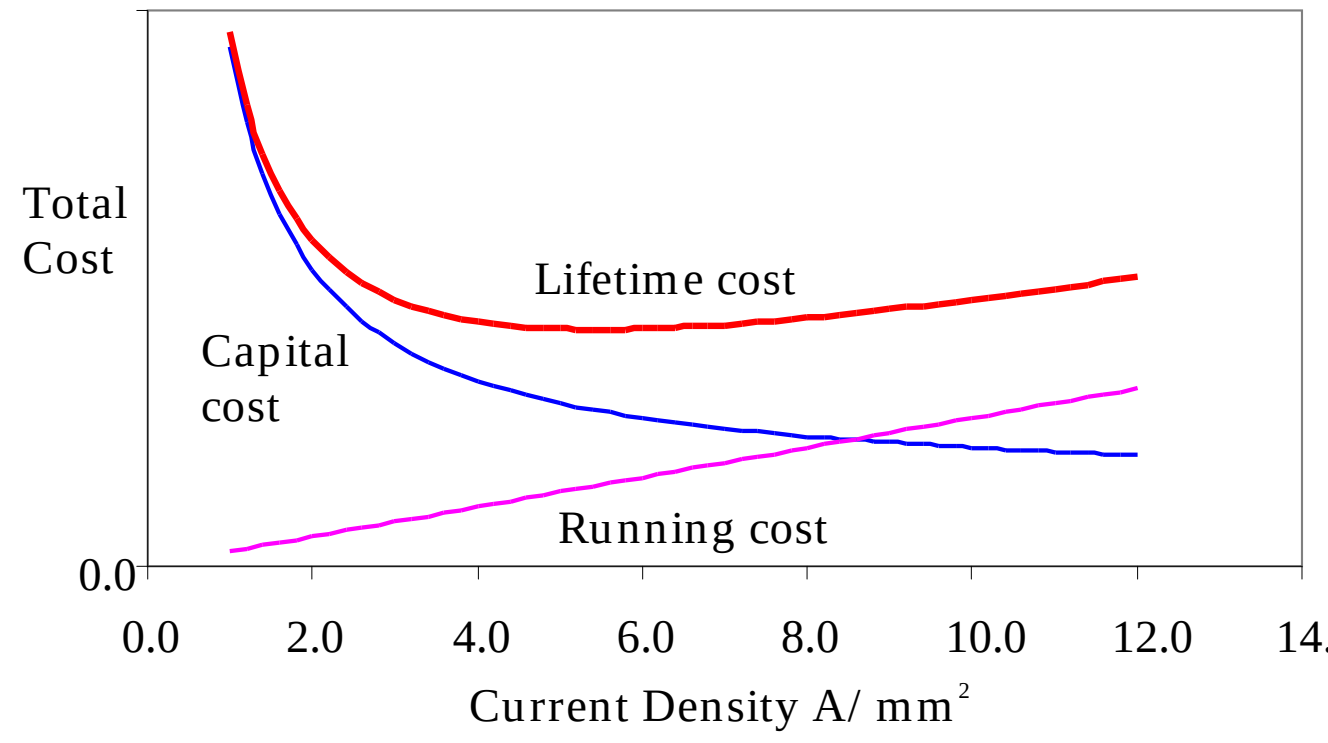

**Fig. 12:** Optimization of current density in the magnet coils, to limit lifetime cost made up of capital and operational components.

In many cases, accelerator project managers choose to design the magnets with current densities below the optimum value, increasing the capital values of the magnet but reducing the temperature rise in the magnets (and decreasing the power loss to make the project more environmentally friendly). This clearly is the route to improve the stability of the magnet.

As an example, the Diamond quadrupoles were designed with a current density, at maximum gradient, of ~2.5 A/mm$^2$. This is above the economic optimum value, which is generally regarded as being in the region of 3.5 to 4 A/mm$^2$. This gave a maximum temperature rise of 10˚C — and perhaps that is a little too high!

## 4.3 Water system induced vibration

This source of instability can really be classified as part of the technical and cultural noise mentioned in Section 4.1.1. However, it is different from other such sources of instability in that it is generated by a system that is part of the accelerator complex and which is, therefore, under the control of those engaged in the accelerator design or operation.

There are two sources of vibration associated with the cooling water that is needed to be circulated through the coils' hollow conductors:

– the mechanical vibration generated by the water pumps and transmitted to the magnets through the water channels; the pumps have rotating systems and, however well balanced these may be, there is always some residual vibration;

– the water passing through the hollow conductor is required to remove heat from the conductor inner tube surface and to do this it must have a velocity which produces turbulent flow; laminar flow does not break the water's boundary layer at the tube walls and therefore does not efficiently remove the dissipated heat; the turbulence generates mechanical vibration.

Figure 13 shows the horizontal spectra of girder vibration with water on and off on a beam-line at the Diamond Light Source [10].

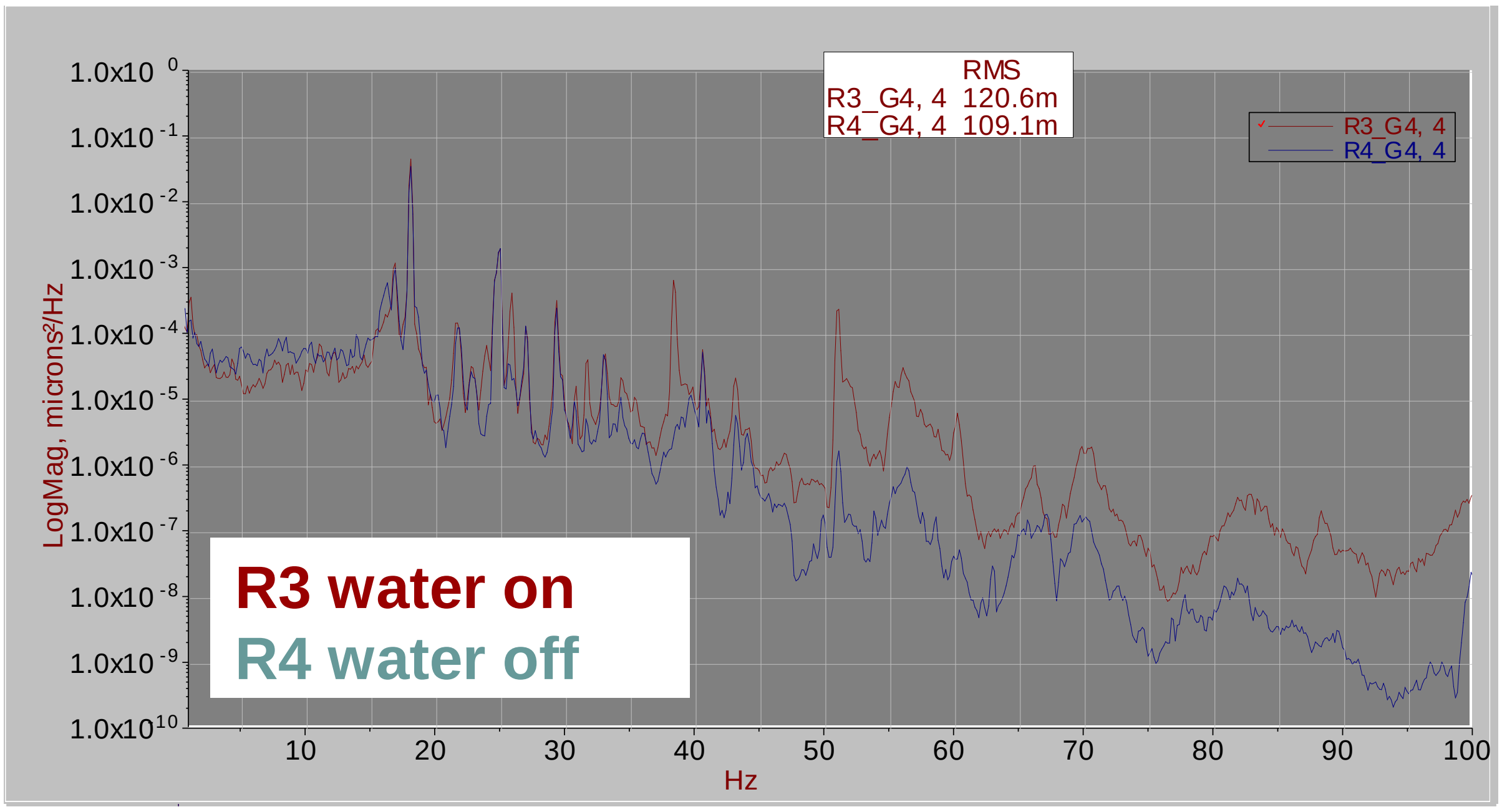

**Fig 13:** Spectra of horizontal girder vibration on a beam-line at the Diamond Light Source [10]

It can be seen that, at low frequencies, there is little difference but above approximately 50 Hz there is greater than an order of magnitude increase in some parts of the spectrum when the water is circulating. But it must be appreciated, however, that these relatively large differences are in regions where the amplitudes are two to three orders of magnitude down on the spectra at low frequencies. They therefore make little contribution to the overall r.m.s. (integrated) values, which are:

R3: $120 \times 10^{-9}$ m;

R4: $109 \times 10^{-9}$ m.

The additional integrated r.m.s. vibration induced by the cooling system is therefore of the order of 10% of the total; not large, but significant and certainly worth minimizing. This therefore also points to minimizing the thermal losses in the coil by designing coils with low current densities that require lower volumes of cooling water and smaller pumps.

## 4.4 Power supply instabilities and ripple

Irrespective of whether the magnets require steady, direct current excitation or an alternating, biased, current waveform, they will be connected to 'power converters' that are fed from the alternating public supply, at an appropriate high voltage, and deliver the necessary power to the magnets, at the right impedance level, with the correct waveform, stabilized, smoothed and controlled. Such operation requires the use of power switching devices and fast feedback servosystems that have a means of controlling the high current and voltage output.

Before the advent of switched power electronics, control was through the use of switching valves (mercury arc rectifiers and thyratrons), which could withstand the high currents and moderate to high voltages. They were, however, relatively slow, commutation usually being possible at a maximum frequency of *c.* 300 Hz. Later, the first solid-state power controlled switches, thyristors, offered similar performance but at reduced cost and with simpler auxiliary apparatus.

Such systems were used to power accelerator magnets up to the 1960s and later. They all generated sharp switching spikes in their output voltage, at their fundamental switching frequency (usually 300 Hz) and higher harmonics. Large smoothing filter circuits were necessary and operational performances with a stability at best of $1{:}10^4$, with similar levels of current ripple were achieved.

The situation is now substantially improved by the invention and commercial availability of far faster power switches, which can be commanded to 'switch off' as well as to become conducting; a facility not available in the older power switching devices. These new devices — insulated gate bi-polar transistors (IGBTs) — can now switch currents well into the thousands of amps at kilovolt potentials, in a few microseconds, though it is necessary to ensure that excessive power dissipation does not occur at the junctions during the switching transients.

These devices are now used in modern 'switch-mode' power converters. A simplified schematic diagram of such a device, to generate a d.c. output, is shown in Fig. 14.

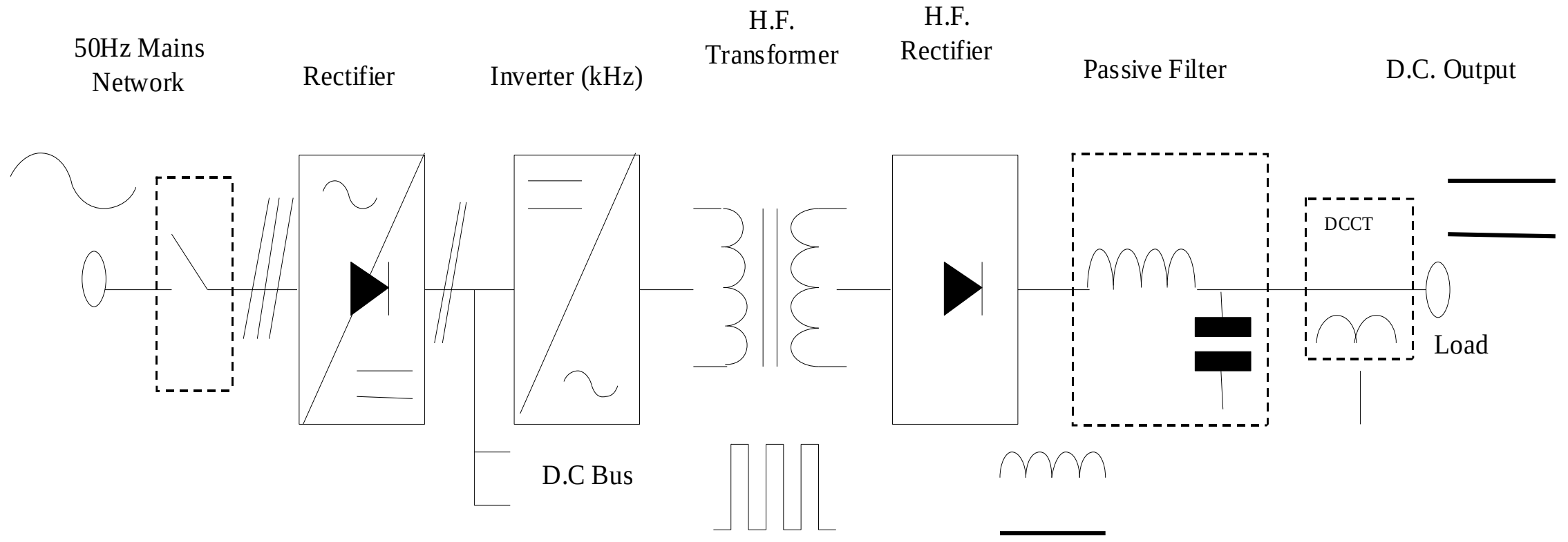

**Fig 14:** Simplified schematic diagram of a typical modern 'switch-mode' power converter

The front-end is powered from the unregulated public supply network. After the usual isolating switch-gear, the power is rectified through a simple unregulated diode system. The output is then fed to a high-frequency inverter, using IGBTs to generate an alternating voltage at frequencies up to 40 kHz (depending on the application and power levels). This is then transformed to meet the output impedance level that is needed for the load. As the size and cost of a power transformer varies roughly as the inverse of its operating frequency, this transformer could be an order of magnitude smaller than one handling the same power at the network supply frequency. The alternating power is again rectified — providing a voltage on the output rail which has a very much higher frequency ripple and which therefore can be smoothed with much smaller filters. Finally, the output current is passed through a direct-current current transformer (DCCT) which provides the signal for an ultra-fast feed-back loop to control the output to a very high stability. This is done through the inverter, which can control its voltage output with a response frequency that is commensurate with its switching time (a few tens of microseconds).

Such systems can provide current stabilities of $1{:}10^5$, nearly as standard, and even of the order of a few ppm at somewhat higher cost; their output ripple is very low (of the same order as their stability) and, in the event of an external fault being detected in the load circuit, they can cut the output power far faster than a circuit-breaker or a fuse; power for power, they are no more expensive than the old power converter systems.

It is clear that they should be the power converters of choice for a modern accelerator magnet system.

## 5. Achieved stability: the current ‘state of the art’

This final section presents the level of stability now being achieved, using the Diamond Light Source as an example. It also points the way to possible further improvement using dynamic beam position control, a technique outside the remit of this paper.

### 5.1 Minimizing the instabilities

Section 4 above gives an overview of the most prevalent sources of instability and gives an indication of the most productive ways of minimizing their effects. These can be briefly summarized as applying the following ‘due diligence’ provisions during project planning and engineering design:

– choose a site with low ground vibration, examining and measuring the ‘technical and cultural’ noise that is present and assessing the impact of the measured spectra on engineering components and, consequentially, the beam;

– design the magnet girders to have high resonant frequencies of *c.*50 Hz or above, obtaining girder response spectra which have such values, well beyond the microseismic peak and in the region where the technical and cultural noise spectrum is also decreasing strongly with frequency;

– design for highly stable temperature control in the accelerator tunnel;

– minimize the waste heat load from auxiliary equipment that is conducted into the accelerator tunnel;

– minimize magnet temperature rise by using low current densities (particularly quadrupoles), adopting a value well below that indicated as being the economic optimum between capital and running cost;

– use as low water velocity as possible whilst still maintaining turbulent flow in the magnet’s cooling water channels, in order to scourge the heat from the conductor inner wall;

– choose water pumps that have minimum mechanical vibration both through their mounts and also transmitted to the water;

– mechanically insulate magnets from water pumps, feed-lines, etc.;

– use best quality, low ripple, highly stable, state-of-the-art power supplies.

## 5.2 Actual performance

Given all the provisions outlined above, what has been achieved in the state-of-the-art synchrotron source Diamond? Figure 15 shows the horizontal and vertical electron beam displacement power spectral density as measured at Diamond [11]. These data were obtained with the feedback orbit control systems (see below) disabled.

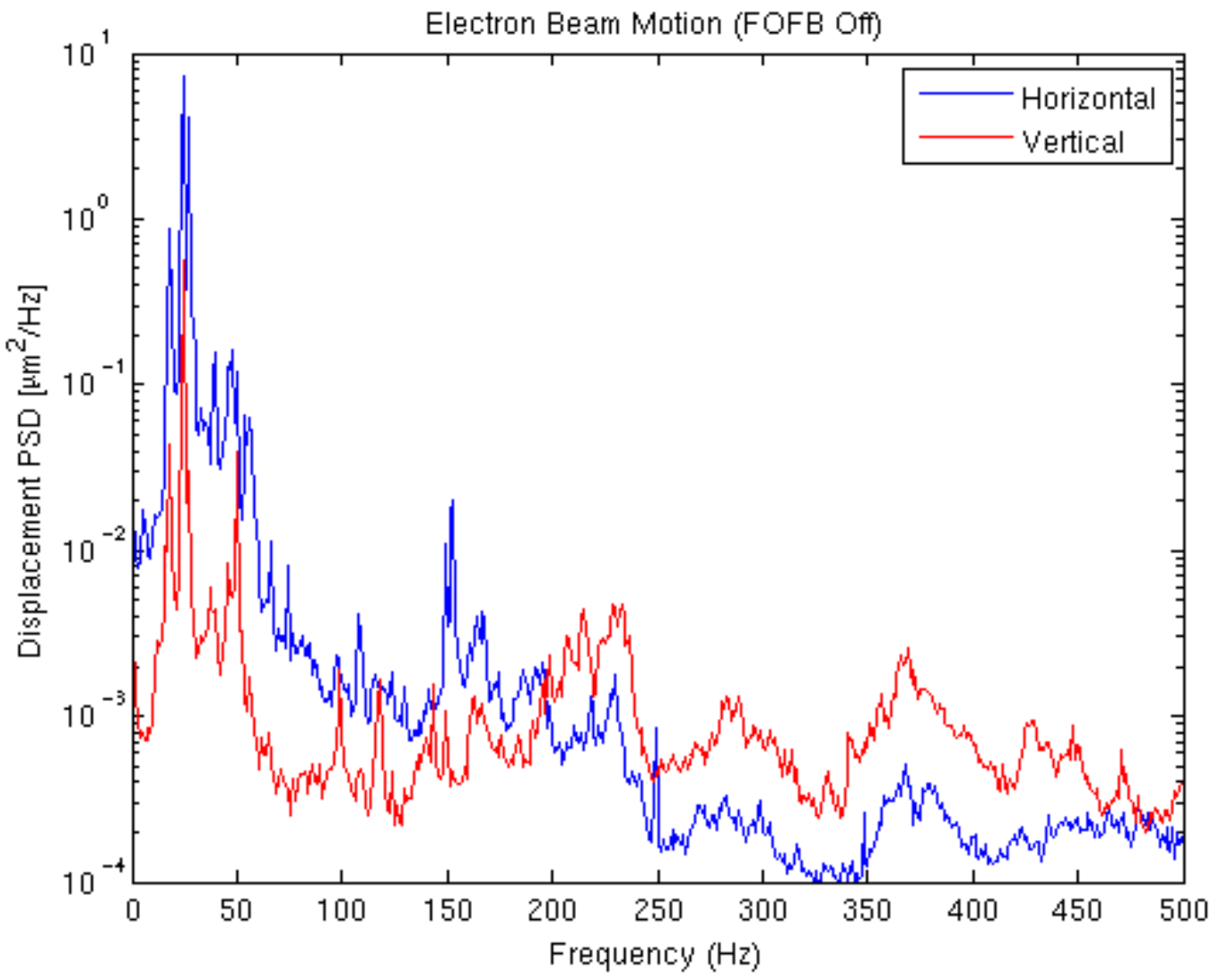

**Fig 15:** The horizontal and vertical electron beam displacement power spectral density measured at the Diamond Light Source with orbit feedback systems disabled

Integrating these data gives the total r.m.s. displacement spectra, as shown in Fig. 16.

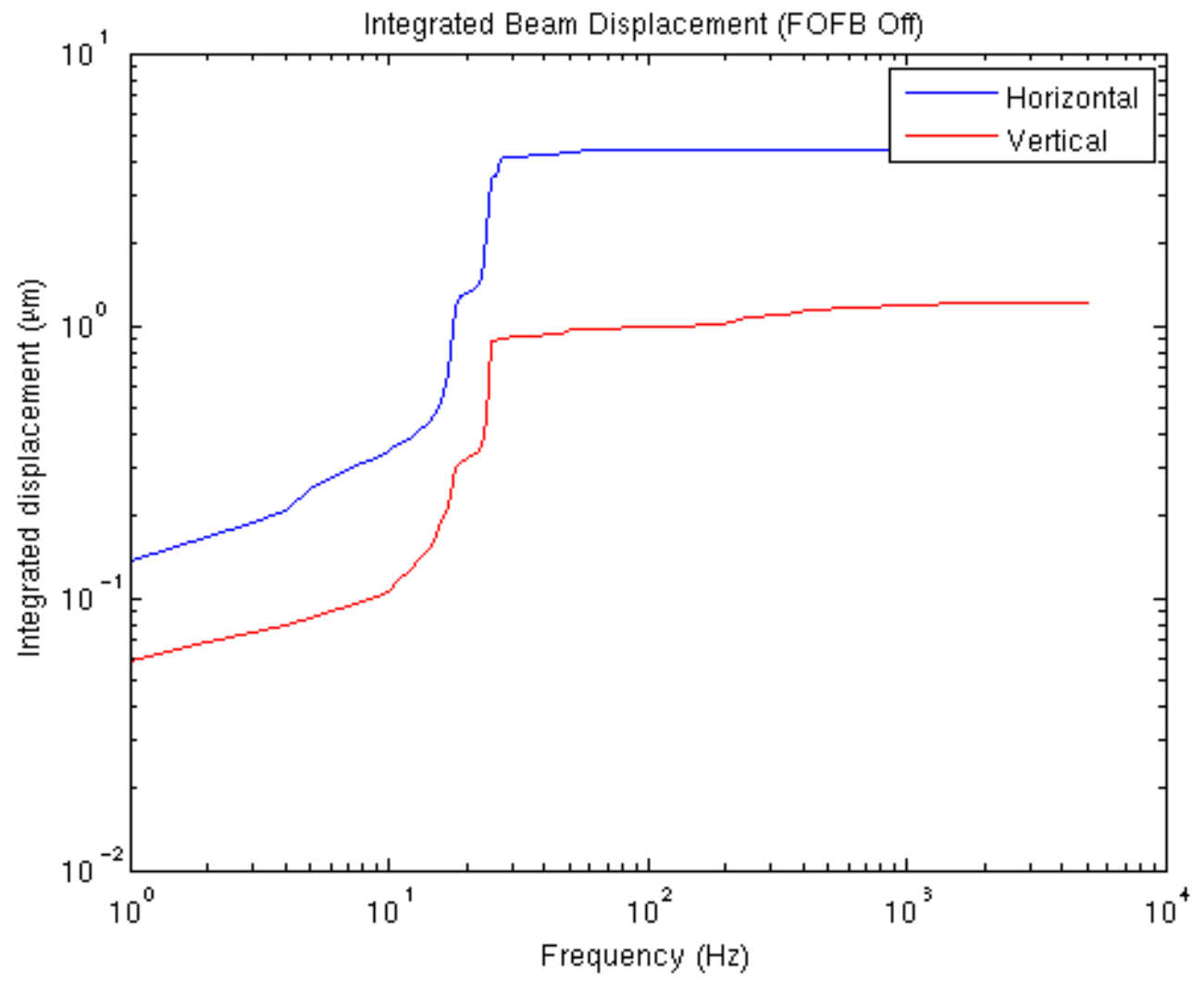

**Fig 16**: The integrated horizontal and vertical electron beam r.m.s displacement, measured at the Diamond Light Source with orbit feedback systems disabled

Note amplitude (r.m.s.) of integrated vibrations:

| | |
|---|---|
| horizontal | 4 µm; with a target value of 12.3 µm; |
| vertical | 1 µm; with a target value of 0.64 µm. |

So the horizontal vibration is well within specification, but the vertical disturbances need to be further reduced by a factor of about 0.5. This indicates that the reduction of vibration achieved by the best available mechanical and electrical engineer provisions is not fully adequate to meet the stringent requirements of beam stability in the accelerator. However, the use of beam-position feedback systems can then give a further reduction in beam motion. This is demonstrated in the figures below, which show  the integrated horizontal and vertical positional motion (Fig. 17) and the horizontal and vertical angular motion (Fig. 18) of the electron beam in 24 straights of the Diamond Light Source, with and without 'fast-orbit feedback' (FOFB).

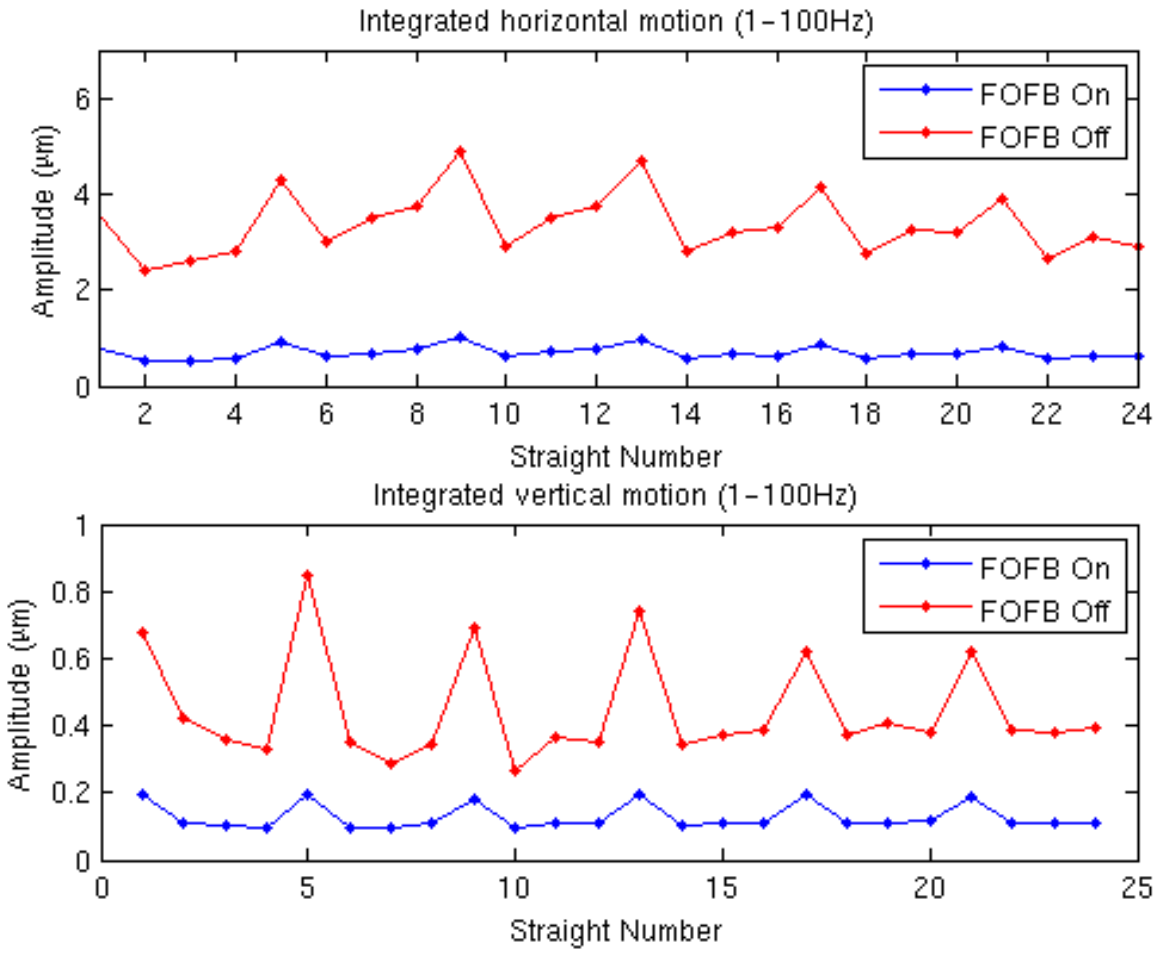

**Fig 17:** The horizontal and vertical positional motion of the electron beam in 24 straights of the Diamond Light Source, with and without 'fast-orbit feedback' (FOFB)

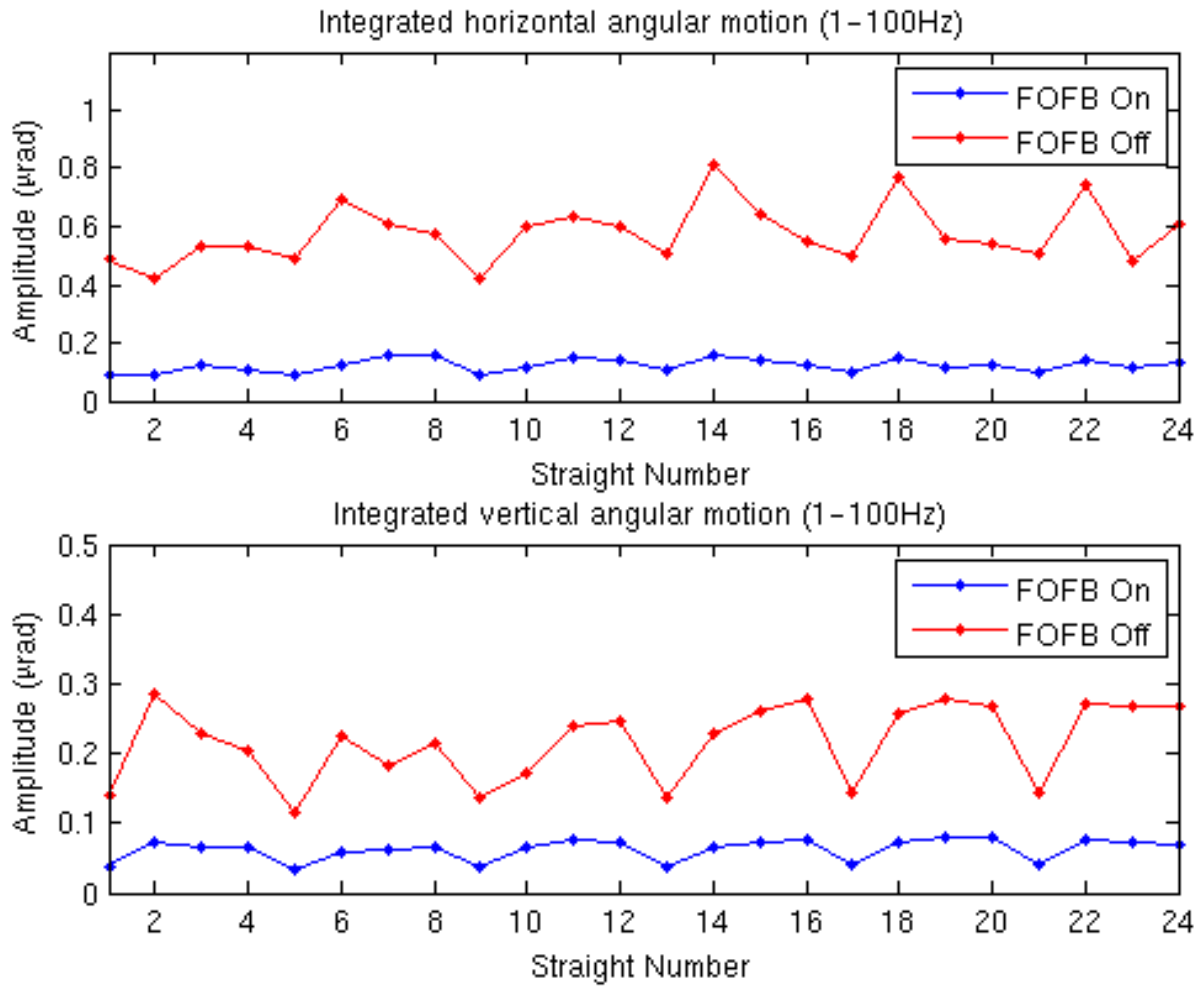

**Fig 18:** The horizontal and vertical angular motion of the electron beam in 24 straights of the Diamond Light Source, with and without 'fast-orbit feedback' (FOFB)

The four sets of data demonstrate clearly that the fast-orbit feedback systems produce substantial improvement in beam stability and thereby meet the original specification in the vertical and horizontal direction. However, it is not the purpose of this paper to explore FOFB systems any further.

## 6 Conclusion

By using best engineering practice and modern techniques for the design, construction, and operation of magnets and their power supplies, beam disturbance due to magnet instability and poor reproducibility can be minimized. But beam-position feedback systems will generally be needed as the final stabilizing influence on the beam and these are now extensively used in light sources, colliders, and other such facilities.

## Acknowledgements

In preparing the presentation and this paper, I made use of much published material — including very new data presented at the Vancouver PAC (June 09) — and also information directly provided by colleagues. I am therefore happy to acknowledge help, either directly or through study of their papers, from Roberto Bartolini, DLS; Nicolas Delerue, University of Oxford; Richard Fielder, DLS; David Holder, Cockcroft Institute and University of Liverpool; Hou-Cheng Huang, DLS; James Kay, DLS; Ian Martin, DLS; Christopher Steier, LBNL; Jörg Wenninger, CERN.